\documentclass[11pt,a4paper,useAMS,usenatbib]{emulateapj}
\usepackage{epsfig}
\usepackage{amsmath}
\usepackage{natbib}
\usepackage{xcolor}

\begin{document}

\title{Constraining the dark matter halo mass \\ of isolated low-surface-brightness galaxies}

\author{Orsolya E. Kov\'acs\altaffilmark{1,2,3}, \'Akos Bogd\'an\altaffilmark{1}, Rebecca E. A. Canning\altaffilmark{4}}
\affil{\altaffilmark{1}Harvard Smithsonian Center for Astrophysics, 60 Garden Street, Cambridge, MA 02138, USA; orsolya.kovacs@cfa.harvard.edu}
\affil{\altaffilmark{2}Konkoly Observatory, MTA CSFK, H-1121 Budapest, Konkoly Thege M. {\'u}t 15-17, Hungary}
\affil{\altaffilmark{3}E{\"o}tv{\"o}s University, Department of Astronomy, Pf. 32, 1518, Budapest, Hungary}
\affil{\altaffilmark{4}Kavli Institute for Particle Astrophysics and Cosmology, Stanford University, 452 Lomita Mall, Stanford, CA 94305-4085, USA}
\shorttitle{LOW-SURFACE-BRIGHTNESS GALAXIES IN X-RAYS}
\shortauthors{KOV\'ACS ET AL.}

\begin{abstract}

Recent advancements in the imaging of low-surface-brightness objects revealed numerous ultra-diffuse galaxies in the local Universe. These peculiar objects are unusually extended and faint: their effective radii are comparable to the Milky Way, but their surface brightnesses are lower than that of dwarf galaxies. Their ambiguous properties motivate two potential formation scenarios: the ``failed'' Milky Way and the dwarf galaxy scenario. In this paper, for the first time, we employ X-ray observations to test these formation scenarios on a sample of isolated, low-surface-brightness galaxies. Since hot gas X-ray luminosities correlate with the dark matter halo mass, ``failed'' Milky Way-type galaxies, which reside in massive dark matter halos, are expected to have significantly higher X-ray luminosities than dwarf galaxies, which reside in low-mass dark matter halos. We perform X-ray photometry on a subset of low-surface-brightness galaxies identified in the Hyper Suprime-Cam Subaru survey, utilizing the \textit{XMM-Newton} XXL North survey. We find that none of the individual galaxies show significant X-ray emission. By co-adding the signal of individual galaxies, the stacked galaxies remain undetected and we set an X-ray luminosity upper limit of ${L_{\rm{0.3-1.2keV}}\leq6.2 \times 10^{37} (d/65 \rm{Mpc})^2 \ \rm{erg \ s^{-1}}}  $ for an average isolated low-surface-brightness galaxy. This upper limit is about {40 times lower} than that expected in a galaxy with a massive dark matter halo, implying that the majority of isolated low-surface-brightness galaxies reside in dwarf-size dark matter halos.
\\
\end{abstract}


\section{Introduction}
\label{sec:intro}

The population of ultra-diffuse galaxies (UDGs) was discovered decades ago in the  Virgo Cluster \citep[e.g.][]{sandage84}. The central surface brightness of these galaxies is lower than that of dwarf galaxies ($\mu_{0}(g)$$\gtrsim$$24 \,\rm{mag \,arcsec^{-2}}$), but their effective radius is comparable to the Milky Way ($r_{\rm{eff}}$$ \gtrsim $$1.5 \,\rm{kpc}$) \citep{2015ApJ...798L..45V}. Recent technological advancements  in the observations of low-surface-brightness systems refocused the interest on UDGs \citep{2014ApJ...782L..24V}. Specifically, the Dragonfly Telephoto Array identified a large population of UDGs in the outskirts of the Coma cluster \citep{dokkum15}, which was followed by the discovery of UDGs in various environments \citep[e.g.][]{2015ApJ...809L..21M,2016AJ....151...96M,2016ApJ...833..168M,2017MNRAS.468.4039R,2017AA...607A..79V,2017MNRAS.467.3751B,2017ApJ...842..133L,2018ApJ...857..104G}. 

Despite the ubiquity of UDGs, their evolutionary path remains ambiguous with two main formation scenarios. UDGs may either be  ``failed'' massive galaxies, which lost their gas at high redshift, thereby preventing any further star formation \citep[e.g.][]{2015ApJ...798L..45V,2016ApJ...828L...6V}. Alternatively, UDGs may be  spatially extended dwarf galaxies, which form by feedback driven gas outflows \citep[e.g.][]{2016ApJ...830...23B,2016MNRAS.459L..51A}. A major difference between these  scenarios is the dark matter halo mass of the galaxies. While the former scenario suggests massive dark matter halos ($M_{\rm{200}} \gtrsim 10^{11} \ \rm{M_{\odot}}$), in the latter scenario, UDGs reside in dwarf-size halos ($M_{\rm{200}} \lesssim 3 \times 10^{10} \ \rm{M_{\odot}}$).
In addition, \citet{2017MNRAS.464L.110Z} suggests that red and blue UDGs may not be a single population, their halo masses could cover a range of masses, and they could be the product of multiple formation scenarios. 

\begin{figure*}[ht]
	\begin{center}
		\leavevmode
		\epsfxsize=1\textwidth \epsfbox{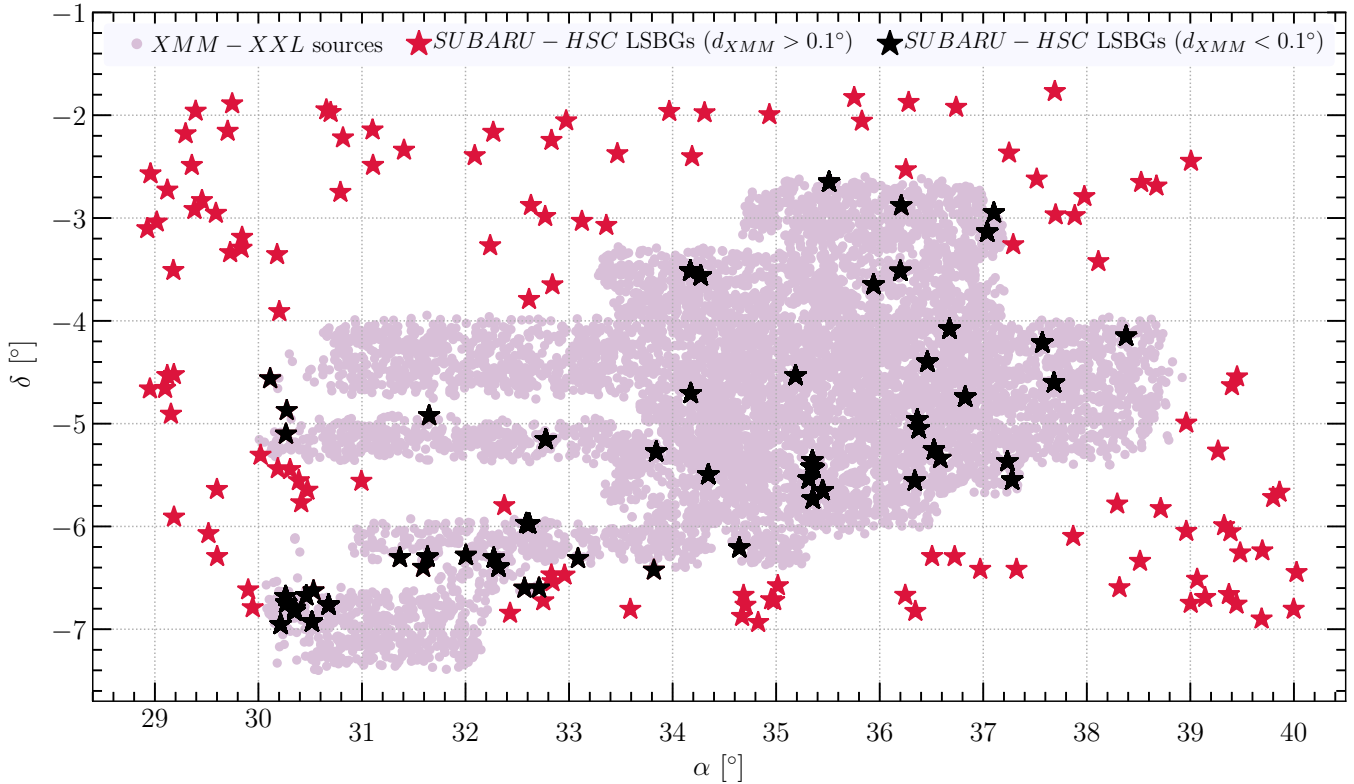}
		\vspace{-.3cm}
		\caption{The sky distribution of a subsample of LSBGs (black and red stars) identified by \citet{2018ApJ...857..104G} in the Subaru HSC-SSP survey overplotted with known sources of the {\textit{XMM-Newton}} XXL North survey (purple points) \citep[]{2016MNRAS.457..110M,2016yCat..74591602L}. We marked LSBGs whose position overlaps with the X-ray field with black stars, and those that are outside the X-ray field with red stars. Sky positions are specified in equatorial coordinates (J2000).}
		\label{fig:skymap}
	\end{center}
\end{figure*}

Since the X-ray luminosity of the hot gas is a robust tracer of the dark matter halo mass \citep{2013ApJ...776..116K,2018ApJ...857...32B}, it is worth exploring the X-ray properties of UDGs. The main contributors of the hot halo gas are the primordial gas accreted at the epoch of galaxy formation, and stellar yields ejected from evolved stars, which then gets heated to the virial temperature of the galaxy \citep{2003ARAA..41..191M,2008MNRAS.388...56B}. If UDGs reside in massive halos, they should retain a significant amount of hot gas. If, however, UDGs preferentially live in dwarf-size halos, they are not likely to have any X-ray emitting gas due to their shallow potential well. Therefore, probing the X-ray gas content of UDGs can constrain their dark matter halo mass and hence their formation mechanism. X-ray observations of field UDGs are demanding with present-day X-ray observatories. 
Searching for UDGs in galaxy clusters may be more promising, but the underlying emission from the intracluster medium may suppress any faint X-ray emission from UDGs. The individual properties of UDGs cannot be probed based on currently available X-ray observations. However, employing a large sample of UDGs and co-adding (i.e.\ stacking) the X-ray emission associated with individual galaxies may provide the sensitivity needed to constrain the X-ray luminosity, and, hence the halo mass of the average UDG population. The advantage of the stacking analysis is that it probes the average UDG population, and removes the distorting effects of outliers and selection effects.

\section{The galaxy sample}
\label{sec:sample}	

A large population of LSBGs was identified in the Hyper Suprime-Cam Subaru Strategic Program (HSC-SSP). Based on the first $200 \ \rm{deg^2}$ of the survey, \citet{2018ApJ...857..104G} detected 781 LSBGs. These galaxies have a wide range of morphologies \citep{2018ApJ...857..104G}, and  a fraction of the sample, depending on the distance of the galaxies, may be UDGs. {It is important to differentiate between LSBGs and UDGs, because these galaxies could have vastly different evolutionary scenarios. While LSBGs are genuine dwarfs, UDGs may be either dwarf or failed massive galaxies.}

Part of the HSC-SSP footprint coincides with the \textit{XMM-Newton} XXL North survey, that maps a region of $25 \ \rm{deg^2}$ using $10$ ks  exposures \citep{2016MNRAS.457..110M}. Combining the optical and X-ray surveys allows a unique way to probe the average X-ray emitting properties of LSBGs. To determine our galaxy sample, we cross-correlated the list of LSBGs identified by \citet{2018ApJ...857..104G} with the footprint of the XXL-North survey. We identified 58 galaxies whose position overlaps with the XXL-North field (Figure \ref{fig:skymap}).
We rejected observations with high background levels, which reduced the final sample size to 51 galaxies (see Appendix). 

The apparent magnitude of the galaxies is $m_{\rm i} = 18-22$\,mag, and the color indices are $g-i = 0.3-1.1 $\,mag, suggesting the presence of both early- and late-type galaxies, in agreement with the general population of LSBGs in the HSC survey. The  $i$-band central surface brightnesses are $\mu_0(i) = 20-26 $ \,mag\,arcsec$^{-2}$. We note that $\sim$$90\%$ of the galaxies have $\mu_0(i) > 23$\,mag\,arcsec$^{-2}$, which is the threshold central surface brightness in the $i$-band for UDGs \citep{2015ApJ...813L..15M,2016ApJS..225...11Y}.  The projected effective radius of the galaxies is in the range of $ r_{\rm{eff}} =3\arcsec-9$\,\arcsec with a mean of $4.6$\,\arcsec. The distance of most galaxies is unknown, although, 6 LSBGs from the full sample of $781$ LSBGs have spectroscopic redshifts and 27 systems are projected to the proximity of a galaxy group \citep{2018ApJ...857..104G}. Based on these, the galaxies likely have a distance of $\sim$$30-100$\,Mpc \citep{2018ApJ...857..104G}, implying mean effective radii of $0.7$\,kpc and $2.2$\,kpc at a distance of $30$\,Mpc and $100$\,Mpc, respectively.

To estimate the stellar mass of the galaxies, we used the $i$-band magnitudes along with the stellar mass-to-light ratio inferred from the $g-i$ color indices \citep{2003ApJS..149..289B}.
The mean stellar mass of the galaxies is $M_{\rm \star} = (0.1 - 1.2) \times 10^{8} \,\rm{M_{\odot}}$ for the distance range of $d=30-100 \,\rm{Mpc}$, implying that our sample consists of galaxies with stellar masses typically observed for dwarf galaxies.

We performed cone searches using the NASA/IPAC Extragalactic Database to infer the environment of the galaxies. We defined a search volume around each galaxy with a radius of $0.5$\,Mpc at the assumed distance of $65$\,Mpc. The specific search radius of $0.5$\,Mpc was chosen considering the virial radius of a typical galaxy group \citep{2015AA...573A.118L}.
We found that 48 of the 51 LSBGs do not have detected neighbors, while 3 LSBGs have $\leq3$ neighbors. Hence, our sample is a homogeneous sample of field LSBGs.

To differentiate between LSBGs and UDG candidates, we rely on $\mu_0(i)$. We define $17$ galaxies, i.e. one-third of the sample with the lowest $i$-band central surface brightness ($\mu_0(i) = 24.5 - 26.3 \rm{\,mag\,arcsec^{-2}}$) as UDG candidates, and $17$ galaxies with the highest $i$-band central surface brightness ($\mu_0(i) = 20.5 - 23.5 \rm{\,mag\,arcsec^{-2}}$) as LSBGs. We also split our sample based on their color and use the $g-i=0.7$ color index to differentiate between blue and red galaxies \citep{2010AA...517A..73G}. This separated our sample to 29 blue and 22 red galaxies. The projected effective radii of the UDG candidate/LSBG and blue/red subsets do not show significant differences between the subsets. 

\section{Analysis of the \textit{XMM-Newton} data}
\label{sec:data}

We collected the raw \textit{XMM-Newton} EPIC-PN data within the XMM-XXL footprint and identified the LSBGs within each observation. 
The list of analyzed observations is listed in the Appendix.

We analyzed the data using the Science Analysis System software package, following \citet{bogdan17,bogdan18}. Specifically, we identified and removed flare contaminated time intervals, using a two step filtering procedure. 
First, we calculated good time intervals (GTI) with 100\,s time bins in the $12-14$\,keV band. Second, we filtered out any remaining flares in the $0.3-10$ keV band with 10\,s binning. We applied $2\sigma$ clipping for both cases. We also excluded the out of time events, and events that are at the border of the CCDs. The original and filtered exposure times are listed in the Appendix.

We created filtered event files and exposure maps in the $0.3-1.2$\,keV band.
We rejected events above $1.2$\,keV, since the characteristic temperature of the hot gas around the LSBGs is not expected to exceed $kT \sim 0.2$\,keV. 

Then, we identified bright point sources, mostly originating from background AGNs.
We performed the source detection on the cleaned images weighted by their respective exposure and PSF maps. Finally, we excluded the detected sources from the images and exposure maps.

\section{Results}
\label{sec:results}

Since the X-ray luminosity of the hot gas is a tracer of the dark matter halo mass \citep{2013ApJ...776..116K,2018ApJ...857...32B}, we measured the $0.3-1.2$\,keV X-ray luminosities for individual galaxies and for the stacked galaxy samples. {Note that these empirical relations were established for early-type galaxies, but they may be applicable for late-type galaxies. Indeed, based on a small sample of nearby systems with stellar masses of $(1-6) \times 10^{11} \ \rm{M_{\odot}}$, \citet{2011MNRAS.418.1901B} established that at the same stellar mass the hot gas content of early- and late-type galaxies are similar.}

We carried out photometry on the individual galaxies. We derived the source counts using a circular aperture with $13\arcsec$ radius, which, at a distance of 65\,Mpc, corresponds to $4$ kpc. To account for the background emission, we used annuli with $60\arcsec-120\arcsec$ radii. We did not obtain statistically significant detections for any of the individual galaxies.

\begin{figure}[t!]
    \leavevmode
      \epsfxsize=0.48\textwidth \epsfbox{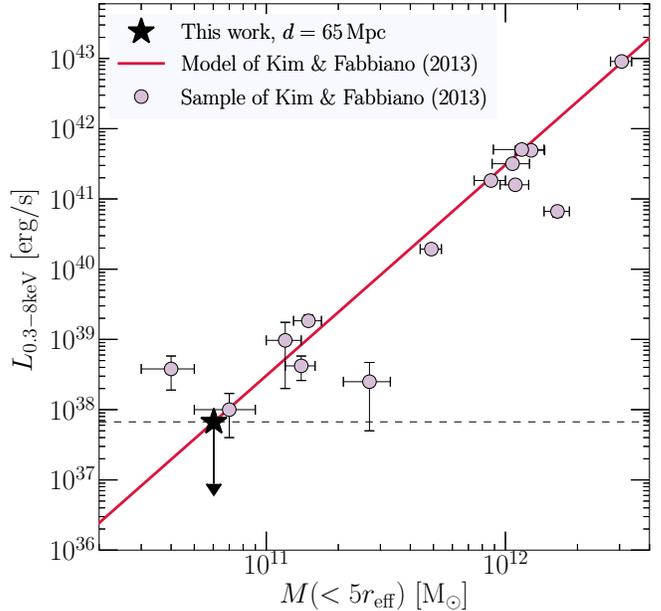}
      \vspace{-.3cm}
      \caption{Empirical relation between the $0.3-8$\,keV band X-ray luminosity of hot diffuse gas and the total gravitating mass within $5 r_{\rm{eff}}$ (red line). We place an upper  limit of ${L_{\rm{0.3-8keV}} \leq 6.4 \times 10^{37} \ \rm{erg \ s^{-1}}}$ on the average X-ray luminosity of field LSBGs assuming a distance of 65 Mpc. The X-ray luminosities of a sample of elliptical galaxies from \citet{2013ApJ...776..116K} is shown with purple points. Galaxies with $M(<5r_{\rm{eff}})\sim 2 \times 10^{11}M_{\odot}$ (or  $M_{\rm 200} = 8 \times 10^{11} \rm{M_\sun}$) are expected to have $L_{\rm{0.3-8keV}} \approx 2.4 \times 10^{39} \ \rm{erg \ s^{-1}}$, which is  $\sim$${40}$ times higher than the  upper limit obtained for the LSBGs of this study.}
     \label{fig:lum}
     \vspace{.3cm}
\end{figure}

\begin{table*}[!t]
\begin{center}
\caption{X-ray Upper limits on the stacked LSBGs}
\begin{minipage}{18cm}
\renewcommand{\arraystretch}{1.3}
\centering
\begin{tabular}{c c c c c c }
\hline 
 & Number of & count rate & $F_{\rm 0.3-1.2keV}$ & $L_{\rm 0.3-1.2keV}$\\
     & galaxies&  [$10^{-4} \ \rm{cts \ s^{-1}}$] & [$10^{-16} \ \rm{erg \ s^{-1} \ cm^{-2}}$] & [$10^{37} \ \rm{erg \ s^{-1}}$ $(d/65 \rm{Mpc})^2]$\\
\hline
Full sample & 51 & $\leq 1.0 $ & $\leq1.3$ & $\leq6.2$  \\
Subsample of UDG candidates & 17 & $\leq1.5 $& $\leq1.9$ & $\leq9.9$ \\
Subsample of LSBGs & 17 & $\leq1.5 $ & $\leq1.9$ & $\leq9.4$ \\
Blue galaxies & 29 & $\leq1.2$ & $\leq1.5$  &  $\leq7.4$ \\
Red galaxies & 22 & $\leq1.7$ & $\leq2.0$  & $\leq10.2$ \\
\hline
\end{tabular} 
\end{minipage}
\end{center}
\label{tab:limit}
\end{table*}  

To increase the signal-to-noise ratio, we co-added the X-ray emission associated with the individual galaxies. For the full sample of 51 LSBGs, this resulted in a total exposure time of $381.1$\,ks. We did not detect statistically significant emission from the stacked data, implying that the X-ray brightness of LSBGs remains below the detection threshold. We also stacked the subset of 17 UDG candidates and the  17 LSBGs as well as the subset of blue and red galaxies in our sample. We did not detect X-ray emission in any of the subsets. 

In the absence of detections, we set upper limits on the average luminosity of galaxies by calculating the confidence bounds of the stacked signal. We defined the $95$\% confidence region of the distribution, and used the upper confidence bound of the count rate as an upper limit.
To convert the count rate to flux, we assumed that the emission from the hot gas can be described with an optically-thin thermal plasma model. Specifically, we assumed a gas temperature of $kT = 0.2$\,keV and a metallicity of $0.2$ solar, which are typical in low-mass elliptical galaxies \citep[e.g.][]{2011ApJ...729...12B,2016ApJ...826..167G}.
Based on the resulting count rate upper limit of $ \leq 1.0 \times 10^{-4}\ \rm{counts \ s^{-1}}$, we estimate the upper limit on the flux of $F_{X}\leq1.3 \times 10^{-16} \ \rm{erg \ s^{-1}  \ cm^{-2}}$ for the full sample of 51 galaxies, which corresponds to a luminosity of $L_{\rm{0.3-1.2keV}}\leq6.2 \times 10^{37} (d/65 \rm{Mpc})^2  \ \rm{erg \ s^{-1}} $, where $d$ represents the average distance of the sample in Mpc. We obtained similar upper limits for the subset of UDG candidates and LSBGs and for the blue and red subsamples, which are listed in Table \ref{tab:limit}.

\section{Discussion}
\label{sec:discussion}

\subsection{Constraining the halo masses}
\label{sec:constraining}

The prototypical UDG, Dragonfly 44, has a virial mass of $M_{\rm 200} = 8 \times 10^{11} \rm{M_\sun}$ \citep{2016ApJ...828L...6V,2017ApJ...844L..11V}. Using its effective radius of $4.6$ kpc and assuming an NFW dark matter halo profile, we estimate that it has a  total mass of $M(<5r_{\rm{eff}})\sim 2 \times 10^{11}M_{\odot}$. Based on the $L_{\rm{X}}-M($$<$$5r_{\rm{eff}})$ scaling relation, this total mass is linked with an X-ray luminosity of $L_{\rm{0.3-8keV}} \approx 2.4 \times10^{39} \ \rm{erg \ s^{-1}}$ in the $0.3 -8$\,keV band. However, the luminosity upper limit measured for the sample of 51 field LSBGs, with an assumed distance of $65$\,Mpc, is ${L_{\rm{0.3-1.2keV}}\leq6.2 \times 10^{37} \ \rm{erg \ s^{-1}}}$ in the $0.3-1.2$\,keV band. After converting the X-ray upper limit luminosity to the $0.3-8$~keV band, we obtain ${L_{\rm{0.3-8keV}} \leq 6.4 \times 10^{37} \ \rm{erg \ s^{-1}}}$. This value is $\sim$${40}$ times lower than that expected for a galaxy with a massive dark matter halo. A similar conclusion is obtained for the subset of 17 UDG candidates and the subset of 17 LSBGs with $0.3-8$~keV as well as for the blue and red galaxy subsets. 

Based on these, we conclude that the dominant population of isolated LSBGs do not host a significant amount of hot X-ray gas, suggesting that they have dwarf-size dark matter halos. Thus, most isolated LSBGs are not ``failed'' Milky Way-type galaxies, but they are likely spatially extended dwarf galaxies. These results are in agreement with semi-analytical and hydrodynamical galaxy formation simulations, which also suggest that UDGs are formed in low-mass dark matter halos \citep{2017MNRAS.470.4231R,2019arXiv190406356L}. {We emphasize that because of the stacking analysis, our results refer to the average galaxy population, hence does not exclude the possibility that a small fraction of galaxies may be luminous and form through the failed Milky Way-scenario.} 

We note that our conclusion is virtually independent of the distance of the galaxies assuming that most of them lie in the distance range of $30-100$\,Mpc \citep{2018ApJ...857..104G}. Indeed, if the average distance of the LSBGs is 100 Mpc, the observed upper limit is {$\sim$16 times lower} than that expected from a galaxy with massive dark matter halo. If, however, the average distance is 30 Mpc, the upper limits are {$\sim$180 times lower} than the X-ray luminosity expected from a galaxy residing in a massive dark matter halo. However, the galaxy sample in this study consists of isolated galaxies. Our results, therefore, may not apply for UDGs in dense environments, such as Dragonfly\,44, for which it is unclear if it is an archetypal or an atypical UDG.

\subsection{Luminosity contribution of X-ray binaries}
\label{sec:luminosity}

The stacked X-ray luminosity measured in the $0.3-1.2$\,keV band is an upper limit on the total X-ray emission associated with the galaxies. The total X-ray luminosity, in addition to the primordial gas emission, includes the emission of the X-ray binary population. Although X-ray binaries are not detected in any of the UDGs in the \textit{XMM-Newton} data, their contribution to the diffuse emission can be estimated considering the correlations between the host galaxy properties and the number of sources. Specifically, low-mass X-ray binaries (LMXBs) are stellar mass tracers, while high-mass X-ray binaries (HMXBs) are star formation rate tracers.
	
Based on the scaling relation obtained for nearby galaxies \citep{2004MNRAS.349..146G} and the average stellar mass of the galaxies assuming a distance of 65\,Mpc (see Section\,\ref{sec:sample}), we estimate that unresolved LMXBs contribute with $9.8 \times 10^{35} \ \rm{erg \ s^{-1}}$ luminosity in the $0.3-1.2$\,keV band. From the simulations of \citet{2017MNRAS.470.4231R}, we assume an average star-formation rate of $9.7 \times 10^{-5}\ \rm{M_{\odot} / yr}$ per galaxy. Based on the scaling relation obtained for star-forming galaxies \citep{2012MNRAS.419.2095M}, we expect $1.1 \times 10^{35} \ \rm{erg \ s^{-1}}$ luminosity from the population of unresolved HMXBs in the $0.3-1.2$\,keV band assuming a distance of 65 Mpc. {These luminosities are significantly lower than the upper limit obtained for the population of LSBGs.}


\subsection{Stellar yields}
\label{sec:stelar}

During their evolution, evolved stars, such as asymptotic giant branch stars, eject a significant amount of gas to their interstellar environment.  After enriched with metals and heated up by supernova explosions, the gas can be observed as diffuse X-ray emission across the galaxy, assuming that the gas is retained in the halo. Considering the contribution of stellar yields is important, because even if the primordial gas was expelled in the early phases of galaxy evolution, stellar yields should be observable at the present epoch.

To compare the luminosity of the gas restored into the interstellar medium with the measured luminosity upper limit, we first calculate the mass originating from stellar yields.
We rely on \citet{1992ApJ...399...76K}, who obtained a mass loss rate of $\sim$$0.0021 \ L_{K}/L_{K,\sun} \ \mathrm{M_{\sun} \ Gyr^{-1}} $ from evolved stars in elliptical galaxies.
Assuming a stellar age of $t_{\rm \star}=10$\,Gyrs and an average $K$-band mass-to-light ratio of $M_{\rm \star}/L_{K} = 0.99$ \citep{2003ApJS..149..289B}, we estimate that in a galaxy with $M_{\rm \star} = 5.1 \times 10^7 \ \rm{M_{\odot}}$, the total mass loss from evolved stars is about $1.1\times10^6 \ \rm{M_{\odot}}$.
We use this mass to calculate the corresponding X-ray luminosity, and assume an XSPEC \texttt{apec} model with gas temperature of $0.2$\,keV, an abundance of $0.2$ solar, and a distance of $65$\,Mpc. In the $0.3-1.2$\,keV band, this gas mass corresponds to a luminosity of $8.6 \times 10^{39} \ \rm{erg \ s^{-1}}$, which is {140 times higher} than the upper limit obtained for LSBGs (Table~1).

The absence of hot gas from stellar yields can be explained by the shallow potential well of LSBGs. In this picture, both supernovae and/or AGN driven winds may be energetic enough to sweep out interstellar medium from the potential well of the galaxies, resulting in negligible X-ray luminosities from the diffuse gas \citep{2006ApJ...653..207D,2008MNRAS.388...56B,2012ApJ...758...65B}. This strengthens  the conclusion that most field LSBGs and UDG candidates reside in low-mass dark matter halos, suggesting that they are not ``failed'' Milky Way type galaxies. 

\begin{center}

{\textsc{appendix}}

\end{center}

We list the analyzed \textit{XMM-Newton} observations, the coordinates, and properties of the the low-surface-brightness galaxies in Table \ref{tab:1}.

\smallskip

\begin{table*}
\caption{List of the analyzed \textit{XMM-Newton} observations and the properties of the low-surface-brightness galaxies.}
\begin{minipage}{18cm}
\renewcommand{\arraystretch}{1.3}
\centering
\begin{tabular}{ccccccccc}
\hline
Observation ID   &   Galaxy ID   & RA        & Dec       & $r_{\mathrm{eff}}$    &   $\mu_{0}(i) $               & $g-i$    &   $t_{\mathrm{orig}}$  &   $t_{\mathrm{clean}}$ \\
                 &               & (J2000)   & (J2000)   & [arcsec]         &   [$\mathrm{mag/arcsec}^2$]   &          &   [ks]                 &   [ks]          \\
(1) & (2) & (3) & (4) & (5) & (6) & (7) & (8) & (9)\\
\hline
0601740201 & 200 & 30.21489$^\circ$  & -6.95317$^\circ$ & 3.94 & 24.23 & 0.46 & 28.66 & 26.57 \\
  & 238 & 30.263$^\circ$  & -6.68032$^\circ$ & 3.77 & 25.29 & 0.81 &   &   \\
0677640140 & 240 & 30.26618$^\circ$  & -5.10157$^\circ$ & 6.19 & 23.08 & 0.56 & 12.75 & 12.11 \\
0601740201 & 241 & 30.26836$^\circ$  & -6.74611$^\circ$ & 4.12 & 24.09 & 0.36 & 28.66 & 26.57 \\
  & 242 & 30.34874$^\circ$  & -6.82286$^\circ$ & 6.96 & 25.28 & 1.03 &   &   \\
0747190132 & 243 & 30.51713$^\circ$  & -6.92962$^\circ$ & 3.45 & 23.99 & 0.90 & 24.43 & 22.79 \\
0677690139 & 259 & 30.67756$^\circ$  & -6.76474$^\circ$ & 9.07 & 24.29 & 0.52 & 9.38 & 8.59 \\
0677670139 & 260 & 31.36609$^\circ$  & -6.30172$^\circ$ & 9.43 & 21.34 & 0.92 & 12.58 & 11.51 \\
0677681234 & 261 & 31.59058$^\circ$  & -6.39757$^\circ$ & 5.72 & 26.31 & 0.81 & 9.57 & 8.68 \\
  & 262 & 31.63289$^\circ$  & -6.29757$^\circ$ & 5.10 & 23.84 & 0.41 &   &   \\
0677630136 & 263 & 31.6484$^\circ$  & -4.92175$^\circ$ & 4.50 & 23.10 & 0.51 & 9.53 & 9.13 \\
0677670137 & 264 & 32.00364$^\circ$  & -6.28043$^\circ$ & 3.74 & 23.82 & 0.52 & 9.44 & 8.83 \\
  & 265 & 32.27289$^\circ$  & -6.30735$^\circ$ & 5.39 & 25.35 & 0.91 &   &   \\
0677680132 & 266 & 32.31839$^\circ$  & -6.39514$^\circ$ & 3.47 & 24.18 & 0.66 & 10.91 & 10.14 \\
0677680131 & 267 & 32.56948$^\circ$  & -6.6003$^\circ$ & 6.27 & 22.65 & 0.40 & 8.83 & 7.80 \\
0677660233 & 268 & 32.58805$^\circ$  & -5.97019$^\circ$ & 4.24 & 20.45 & 0.62 & 9.53 & 8.64 \\
  & 269 & 32.61469$^\circ$  & -5.97442$^\circ$ & 3.57 & 22.37 & 1.13 &   &   \\
0677680131 & 271 & 32.70979$^\circ$  & -6.59894$^\circ$ & 4.40 & 23.87 & 0.80 & 8.83 & 7.80 \\
0677640132 & 304 & 32.7751$^\circ$  & -5.15336$^\circ$ & 3.08 & 24.27 & 0.77 & 10.97 & 10.33 \\
0677670134 & 306 & 33.08616$^\circ$  & -6.31078$^\circ$ & 6.78 & 23.53 & 0.93 & 9.40 & 8.49 \\
0655343859 & 307 & 33.81903$^\circ$  & -6.42413$^\circ$ & 3.51 & 25.15 & 0.58 & 2.40 & 2.27 \\
0600090401 & 308 & 33.84203$^\circ$  & -5.27461$^\circ$ & 4.37 & 22.59 & 0.56 & 50.76 & 47.69 \\
0404968401 & 309 & 34.17093$^\circ$  & -3.51542$^\circ$ & 4.39 & 25.30 & 1.06 & 9.13 & 8.40 \\
0112371701 & 314 & 34.17509$^\circ$  & -4.70562$^\circ$ & 3.95 & 23.50 & 0.90 & 22.61 & 21.07 \\
0404968401 & 315 & 34.27163$^\circ$  & -3.56117$^\circ$ & 3.13 & 25.78 & 0.95 & 9.13 & 8.40 \\
0112370701 & 316 & 34.34471$^\circ$  & -5.49559$^\circ$ & 3.05 & 24.14 & 0.44 & 42.92 & 41.77 \\
0747190837 & 317 & 34.64441$^\circ$  & -6.20896$^\circ$ & 8.67 & 25.02 & 0.67 & 11.65 & 11.08 \\
0785100601 & 318 & 35.18785$^\circ$  & -4.53235$^\circ$ & 7.75 & 23.14 & 0.59 & 21.97 & 20.19 \\
0785101001 & 319 & 35.3154$^\circ$  & -5.53917$^\circ$ & 3.81 & 23.23 & 0.77 & 17.13 & 16.28 \\
  & 321 & 35.34326$^\circ$  & -5.42328$^\circ$ & 3.93 & 24.59 & 0.41 &   &   \\
  & 322 & 35.3498$^\circ$  & -5.35893$^\circ$ & 3.90 & 23.36 & 0.38 &   &   \\
0553910701 & 323 & 35.35374$^\circ$  & -5.73281$^\circ$ & 4.86 & 23.15 & 0.48 & 11.74 & 10.85 \\
0785101001 & 327 & 35.36102$^\circ$  & -5.43343$^\circ$ & 4.74 & 25.38 & 0.62 & 17.13 & 16.28 \\
0785101601 & 328 & 35.44907$^\circ$  & -5.65036$^\circ$ & 4.08 & 24.53 & 0.95 & 16.65 & 16.07 \\
0147110101 & 330 & 35.51138$^\circ$  & -2.65082$^\circ$ & 3.05 & 25.90 & 0.84 & 9.03 & 8.42 \\
0037980201 & 332 & 35.94019$^\circ$  & -3.64859$^\circ$ & 3.59 & 23.92 & 0.63 & 7.73 & 7.13 \\
0037980801 & 333 & 36.19941$^\circ$  & -3.51751$^\circ$ & 4.83 & 24.35 & 0.31 & 8.82 & 8.10 \\
0785102301 & 334 & 36.33909$^\circ$  & -5.5587$^\circ$ & 3.89 & 25.33 & 0.33 & 16.75 & 15.68 \\
0780450901 & 337 & 36.3633$^\circ$  & -4.96164$^\circ$ & 3.22 & 24.98 & 0.50 & 14.73 & 13.37 \\
0111110301 & 338 & 36.37673$^\circ$  & -5.04815$^\circ$ & 3.98 & 25.57 & 0.50 & 17.74 & 16.74 \\
0780450701 & 340 & 36.46018$^\circ$  & -4.4021$^\circ$ & 3.48 & 23.98 & 1.01 & 13.12 & 12.52 \\
0780451201 & 342 & 36.52391$^\circ$  & -5.25631$^\circ$ & 3.75 & 24.82 & 0.32 & 21.80 & 20.70 \\
  & 343 & 36.58562$^\circ$  & -5.33722$^\circ$ & 3.13 & 23.48 & 0.58 &   &   \\
0780451601 & 344 & 36.67423$^\circ$  & -4.07979$^\circ$ & 5.46 & 25.46 & 1.02 & 28.73 & 26.88 \\
0109520201 & 346 & 36.82575$^\circ$  & -4.74147$^\circ$ & 4.13 & 24.39 & 0.46 & 18.61 & 17.88 \\
0037981501 & 351 & 37.10237$^\circ$  & -2.94921$^\circ$ & 4.54 & 24.28 & 0.84 & 8.81 & 8.44 \\
0404964701 & 352 & 37.23321$^\circ$  & -5.36786$^\circ$ & 3.50 & 23.49 & 0.94 & 8.73 & 8.58 \\
0677600134 & 353 & 37.2788$^\circ$  & -5.54917$^\circ$ & 5.05 & 23.39 & 0.84 & 9.27 & 8.64 \\
0677580133 & 354 & 37.57071$^\circ$  & -4.21543$^\circ$ & 3.98 & 23.66 & 0.61 & 11.47 & 10.58 \\
0677580134 & 357 & 37.68303$^\circ$  & -4.60199$^\circ$ & 4.34 & 24.23 & 0.92 & 9.48 & 8.89 \\
0677580139 & 360 & 38.37947$^\circ$  & -4.14898$^\circ$ & 3.57 & 23.06 & 0.67 & 11.07 & 10.00 \\
\hline
\end{tabular} 
\vspace{0.1in}
\end{minipage}
Columns are as follows: (1) \textit{XMM-Newton} identifier of the analyzed observation; (2) unique galaxy identifier taken from \citet{2018ApJ...857..104G}; (3) and (4) coordinates of galaxies in decimal degrees (J2000) \citep{2018ApJ...857..104G}; (5) projected effective radii; (6) $i$-band central surface brightness; (7) $g-i$ color indices of galaxies \citep{2018ApJ...857..104G}; (8) and (9) original and clean PN exposure times.
\label{tab:1}
\end{table*}

\smallskip

\begin{small}
\noindent
\textit{Acknowledgements.}
{We thank the referee for the constructive report.} This work uses observations obtained with \textit{XMM-Newton}, an ESA science mission with instruments and contributions directly funded by ESA Member States and NASA. This research has made use of the NASA/IPAC Extragalactic Database (NED), which is operated by the Jet Propulsion Laboratory, California Institute of Technology, under contract with the National Aeronautics and Space Administration. \'A.B. acknowledges support from the Smithsonian Institution.
\end{small}

\end{document}